\documentclass[11pt]{article}
\usepackage[margin=1in]{geometry}
\usepackage{natbib}
\usepackage{amssymb}

\providecommand{\pagerange}[1]{}
\providecommand{\pubyear}[1]{}
\providecommand{\volume}[1]{}
\providecommand{\artmonth}[1]{}
\providecommand{\doi}[1]{}
\providecommand{\backmatter}{}
\newenvironment{keywords}{\par\medskip\noindent\textbf{Key words:} }{\par\medskip}
\date{}

\usepackage[figuresright]{rotating}
\usepackage{amsmath}
\usepackage[hyphens]{url}
\usepackage{hyperref}
\usepackage[utf8]{inputenc}
\usepackage{graphicx}
\usepackage{longtable}
\usepackage{booktabs}
\usepackage{threeparttable}
\usepackage[section]{placeins}% keep each float inside the section that describes it
\usepackage{multirow}

\newcommand{\dstar}{\delta^{\ast}}
\newcommand{\Chat}{\widehat C}
\newcommand{\shat}{\hat s}
\newcommand{\Dhat}{\widehat\Delta}
\newcommand{\argmax}{\operatorname*{arg\,max}}

\title{Does a Developed Comorbidity Index Really Add Value? A
Selection-Aware Bootstrap for Post-Selection Concordance}

\usepackage{array}
\usepackage{here}
\usepackage{lscape}
\author{M. Ehsan Karim\\[4pt]
\normalsize
\begin{minipage}{0.82\textwidth}\centering
School of Population and Public Health, University of British Columbia, Vancouver, British Columbia, Canada; and Centre for Advancing Health Outcomes, St.\ Paul's Hospital, Vancouver, British Columbia, Canada\\[3pt]
\texttt{ehsan.karim@ubc.ca}
\end{minipage}}

\begin{document}

\date{}
\pagerange{\pageref{firstpage}--\pageref{lastpage}} \pubyear{}
\volume{0}
\artmonth{January}
\doi{0000-0000-0000}
\label{firstpage}

\maketitle

\begin{abstract}
Disease-specific comorbidity indices are routinely developed by building
several candidate constructions and reporting the best-scoring one, then
claiming it adds discriminative value over a fixed off-the-shelf comparator
such as the Charlson or Elixhauser score. We show that the optimism
correction in standard use does not make that claim valid. Because it
corrects the selected model as if it were the only one ever fit, it omits
the winner's-curse term from choosing the best of several candidates;
so its confidence interval for the incremental concordance is not
merely optimistic in small samples but structurally miscalibrated, and
does not shrink as the sample grows. At a true null it
inflates false claims of added value above the nominal
level, increasingly so as more candidates are screened. We introduce a
drop-in selection-aware bootstrap that re-runs the best-of-several selection
inside each resample with the comparator held fixed, removing the
structural bias. In a fully-known-truth simulation, 95\% coverage
under the standard correction falls
from 0.94 with one candidate to 0.70 with a hundred, while the
selection-aware interval holds near nominal; its coverage matches
a calibrated cross-validation interval, and at a matched error rate
it is at least as powerful. The results hold under Uno's concordance, and a
semi-synthetic experiment on real survey data confirms when the correction matters. In
practice, if several constructions were tried, report a selection-aware
interval, most needed with many similar-quality candidates and few events per candidate.
The scope is
discrimination only; software and results reproduce every
finding.
\end{abstract}

\begin{keywords}
Comorbidity index; Concordance; Optimism correction;
Post-selection inference; Winner's curse.
\end{keywords}

\section{Introduction}\label{introduction}

\textbf{Background and Motivation}: Comorbidity indices are a staple of
prognostic and risk-adjustment modelling. When a condition-specific cohort
is available, a common and reasonable ambition is to build a
disease-tailored index that captures prognostically relevant comorbidity
better than a generic, off-the-shelf instrument such as the Charlson or
Elixhauser score \citep{Charlson1987, Elixhauser1998}. In practice,
``building the index'' is rarely a single decision. A team assembles a set
of candidate comorbidity flags, then explores several
constructions---different flag subsets, coding and grouping choices,
weighting schemes, or variable-selection rules---fits each, and reports the
one with the best apparent performance on the development sample. This
best-of-$K$ workflow is normal, sensible, and almost never described as a
source of bias.

Yet reporting the maximum of $K$ noisy performance estimates is exactly
the setting in which the \emph{winner's curse} operates. The selected
index looks good partly because it is genuinely good and partly because it
was the luckiest of the candidates tried on this particular sample; the
more constructions are sifted, the larger the share of the apparent edge
that is luck rather than merit. The apparent incremental value of the
developed index over the comparator is therefore optimistically inflated,
and the inflation grows with the number of candidates. Because the
confidence interval for that incremental value drives the central claim of
most applied comorbidity-index papers---``our index adds value over
Charlson/Elixhauser''---an interval that is too narrow translates directly
into systematic over-claiming.

\textbf{The estimand}: We make the target of inference explicit. Let
$C(\cdot)$ denote Harrell's concordance for a right-censored survival
outcome. The quantity of interest is the incremental discrimination
\begin{equation}
\Delta \;=\; C(\text{best-of-}K\text{ developed index}) \;-\; C(\text{fixed comparator}),
\label{eq:estimand}
\end{equation}
the difference between the concordance of the data-developed, selected
index and that of a fixed off-the-shelf comparator applied as-is. Two
features distinguish this estimand. First, it is \emph{asymmetric by
design}: the developed arm is chosen and fit on the data (and is therefore
optimistic), whereas the comparator is a frozen formula that is never
re-fit (and carries no optimism). Second, it is a \emph{post-selection,
conditional} target---the truth for the model actually selected in a given
analysis---because that is the object about which applied teams make their
claim. Valid inference on $\Delta$ must honour both the asymmetry and the
selection-over-$K$-constructions step. Throughout, an interval's \emph{coverage} is how
often it actually contains the true value: a valid $95\%$ interval should cover $95\%$ of
the time, and one that \emph{under-covers} is too narrow and over-states the evidence for
added value. We report Harrell's $C$ as the
primary functional for continuity with practice; because its
censoring-dependent shift is score-specific and need not cancel in
$\Delta$, the confirmatory analysis uses Uno's
inverse-probability-of-censoring-weighted (IPCW) $C$ \citep{Uno2011}
(Section~\ref{setup}). The selection mechanism we characterize is
orthogonal to this choice.

\textbf{Position relative to prior literature}: \emph{(Readers who want the recommendation
may skip to \textbf{Contributions} below; Table~\ref{tab:gap} summarizes where this paper
sits.)} The building blocks we rely on
are established, and we claim none of them as new. Optimism correction for
a \emph{single} prediction model---via the bootstrap and the
.632/.632\texttt{+} constructions \citep{Efron1983, EfronTibshirani1997}---is
standard \citep{EfronGong1983, Harrell2015, Steyerberg2001}, and
location-shifted and two-stage bootstrap intervals for such
optimism-corrected accuracy measures are given by \citet{Noma2021}. Valid error estimation under model selection through nested or repeated
cross-validation is likewise well developed, from \citet{VarmaSimon2006} to
the recent nested cross-validation variance of
\citet{BatesHastieTibshirani2024} and its survival-outcome extension
\citep{SunTibshirani2023}. The broader idea that selecting a winner biases
subsequent inference---the minimum-error cross-validation bias of
\citet{TibshiraniTibshirani2009}, the order-statistic view of
selection-induced bias \citep{McLatchieVehtari2024}, and the selective- and
post-selection-inference literature \citep{Berk2013, AKM2024}---is well
studied; closest to us, \citet{RinkBrannath2024} give valid confidence
bounds for the performance of a model \emph{selected} from several
candidates. Comorbidity-index \emph{construction} itself---mapping
coefficients to integer points \citep{Sullivan2004, vanWalraven2009},
stability selection \citep{MeinshausenBuhlmann2010}, calibration---is mature
and is not a methodological contribution here.

What none of these addresses directly is the specific estimand of
Equation~(\ref{eq:estimand}). Single-model optimism correction solves a
different problem: it corrects the fitted winner as if it were the only
model ever considered, and so silently omits the selection component of
the optimism. Generic nested cross-validation targets a
selection-\emph{averaged} error, not the conditional incremental value of
the selected model over a fixed reference, and does not speak to the
asymmetry between a re-fit arm and a frozen arm. The classical remedy of
bootstrapping the \emph{entire} model-building process
\citep{EfronGong1983, Steyerberg2001, Harrell2015} is precisely the principle
our selection-aware interval applies to the best-of-$K$ argmax---we claim it
as neither new nor our contribution; what in-house practice actually does is
correct the \emph{selected} winner only, omitting the across-construction
selection, and it is that omission we name and quantify. Even
\citet{RinkBrannath2024}, the closest prior work, bounds the performance of
the selected model on its own rather than the \emph{asymmetric difference}
against a never-refit comparator that our estimand targets. The gap we fill
is coverage-valid interval inference for \emph{this} estimand---an
asymmetric, paired difference of two concordances with a built-in
selection-over-$K$ step---together with a usable account of when it
matters.

\begin{table}[htbp]\centering
\caption{Where this paper sits. Each prior ingredient is established and valid for its
own target; none delivers a coverage-valid interval for the asymmetric
``best-of-several developed index minus fixed comparator'' concordance difference of
Equation~(\ref{eq:estimand}).}
\label{tab:gap}
\small
\setlength{\tabcolsep}{5pt}
\begin{tabular}{@{}p{0.27\textwidth}p{0.29\textwidth}p{0.35\textwidth}@{}}
\toprule
Prior approach & What it provides & What it leaves open for our estimand \\
\midrule
Single-model optimism correction (bootstrap, .632$+$; \citealp{Efron1983,Noma2021})
& A valid interval for one pre-specified model
& Corrects the winner as if it were the only model, omitting the across-candidate
selection term \\
Nested / repeated cross-validation
(\citealp{BatesHastieTibshirani2024,SunTibshirani2023})
& Valid error for the whole selection procedure
& Targets selection-\emph{averaged} (marginal) error, not the conditional value of the
reported index; ignores the re-fit-versus-frozen asymmetry \\
Post-selection inference for the winner (\citealp{RinkBrannath2024,Berk2013,AKM2024})
& A valid bound on the selected model's own performance
& Bounds one arm alone, not the asymmetric \emph{paired difference} against a
never-refit comparator \\
Comorbidity-index construction
(\citealp{Sullivan2004,vanWalraven2009,MeinshausenBuhlmann2010})
& Mature machinery for building indices
& Not an inference method for added value \\
\midrule
\textbf{This paper} & \textbf{A drop-in selection-aware bootstrap}
& \textbf{Coverage-valid interval for this estimand, plus a map of when the correction
matters} \\
\bottomrule
\end{tabular}
\end{table}

\textbf{Contributions}: The paper has two parts. First, we show---and confirm by
simulation---that the near-universal in-house practice (correct the selected model's
optimism as if it were the only one fit, then compare to the fixed comparator)
\emph{under-covers}: its $95\%$ interval is too narrow, and the shortfall worsens as more
candidates are tried, with coverage falling from about $0.94$ at $K{=}1$ to $0.70$ at
$K{=}100$ in our headline regime. The reason is a missing \emph{selection} term---the
winner's curse of taking the best of $K$ correlated estimates. Crucially, this error
\emph{does not shrink as the sample grows}, so a larger cohort will not fix it; at a true
null it inflates the rate of false ``adds value'' claims to about $30\%$ at $K{=}100$
(nominal $2.5\%$).

Second, a drop-in fix: folding the ``pick the best of $K$'' step into the bootstrap while
holding the comparator fixed---the \emph{selection-aware bootstrap}---restores near-nominal,
$K$-stable coverage, about $0.92$ across $K$. We validate that near-$0.92$ level in
Section~\ref{simulation} against a bracket of cross-validation intervals spanning
known-loose to known-tight, and show the selection-aware interval matches the calibrated one
and is at least as powerful at a matched error rate. We also map \emph{when} the problem
bites---large $K$, similar-quality candidates, and small events-per-variable (few outcome
events per candidate flag) \citep{Riley2019}---so practitioners know when the correction is
and is not material.

\textbf{Scope of the estimand.} The estimand is
\emph{discrimination only}: a concordance difference is itself an
insensitive measure of a marker's added value \citep{Pencina2008}, and we do
not treat calibration or net benefit. And one theoretical point is left
\emph{open}: at exact candidate ties the ordinary $n$-out-of-$n$ bootstrap
of an argmax is delicate \citep{Andrews2000, FangSantos2019}, and a
complete consistency proof there---with the explicit $m$-out-of-$n$
subsampling rate---is left open.

\textbf{Roadmap}: Section~\ref{setup} fixes the setup and the estimand.
Section~\ref{methods} defines the six interval procedures compared (naive,
single-step, selection-aware, and three nested cross-validation brackets).
Section~\ref{theory} states the mechanism as four propositions.
Section~\ref{simulation} describes the simulation design and reports the
main results---coverage, boundary Type-I error, the structural
non-vanishing (persistence) signature, and where the fix sits on the
power/Type-I frontier. Section~\ref{application} leads with a real-NHANES
coverage confirmation of the mechanism and routes the (modest) decision-impact
illustration to Web Appendix~D. Section~\ref{discussion} discusses scope, open
theoretical questions, and practical guidance. Full statements, proofs, and
complete results are in the Web Appendices; open-source software reproduces
every number. Table~\ref{tab:roadmap} gives the paper at a glance---the
questions we address, our answers and recommendations, and where each is
established.

\begin{table}[htbp]\centering\small
\caption{Roadmap of the paper: the questions we address, our answers and
recommendations, and where each is established. The two central tables are the
methods comparison (Table~\ref{tab:methods}) and the real-data confirmation
(Table~\ref{tab:nhanes}).}
\label{tab:roadmap}
\setlength{\tabcolsep}{5pt}
\begin{tabular}{@{}p{0.22\textwidth} p{0.60\textwidth} p{0.12\textwidth}@{}}
\toprule
Question / decision & Question addressed, and our answer or recommendation & Where \\
\midrule
\multicolumn{3}{@{}l}{\emph{Method and mechanism}}\\
\addlinespace[2pt]
Which interval to report &
\emph{When several index constructions were tried, which interval should you
report?} Fold the best-of-$K$ selection into the bootstrap and hold the
comparator fixed (\emph{selection-aware}); the naive and single-step intervals
under-cover. &
Table~\ref{tab:methods}; \S\ref{methods} \\
\addlinespace
Why single-step fails &
\emph{Why does the standard single-step correction under-cover, and will more
data fix it?} It omits the winner's-curse selection term, so its bias-to-standard-error
ratio stays $O(1)$ and does not shrink as $n$ grows---the failure is structural. &
\S\ref{theory}; Table~\ref{tab:dstar} \\
\addlinespace
Why not cross-validate &
\emph{Why not simply cross-validate instead?} Cross-validation targets the
marginal value of the whole selection procedure; the selection-aware bootstrap
targets the conditional value of the one index actually reported. &
\S\ref{theory}; Web App.~C \\
\addlinespace
\multicolumn{3}{@{}l}{\emph{Evidence}}\\
\addlinespace[2pt]
How bad it gets &
\emph{How much does coverage degrade as more candidates are screened?} 95\%
coverage falls from 0.94 at $K{=}1$ to 0.70 at $K{=}100$ under single-step,
while selection-aware holds near 0.92. &
Table~\ref{tab:coverage}; Fig.~\ref{fig:coverage} \\
\addlinespace
Whether the fix is valid &
\emph{Is the selection-aware interval genuinely valid, not just better than a
weak baseline?} Its coverage matches a calibrated cross-validation
interval, and at a matched error rate it is at least as powerful. &
Table~\ref{tab:frontier} \\
\addlinespace
Real data &
\emph{Does the same failure appear on real data?} On NHANES
comorbidity--mortality covariance the near-exchangeable regime reproduces the
$O(1)$ bias, and selection-aware repairs coverage. &
Table~\ref{tab:nhanes}; \S\ref{application} \\
\addlinespace
\multicolumn{3}{@{}l}{\emph{Practice}}\\
\addlinespace[2pt]
Compute cost &
\emph{Does the fix cost more compute?} No---the three bootstrap intervals share
one resampling loop; only the block jackknife is pricier ($\sim2\times$). &
Table~\ref{tab:methods} \\
\addlinespace
When it matters &
\emph{When is the correction material?} Large $K$, similar-quality candidates,
and small events-per-variable; a rule-of-thumb guide is provided. &
Table~\ref{tab:rot}; \S\ref{discussion} \\
\bottomrule
\end{tabular}
\end{table}

\section{Setup and Estimand}\label{setup}

\subsection{Development sample and candidate library}\label{setup-library}

We observe a development sample of $n$ independent subjects, each with a
vector of binary comorbidity flags $X\in\{0,1\}^p$ and a right-censored
survival outcome $(\tilde T,\delta)$, $\tilde T=\min(T,\mathcal C)$,
$\delta=\mathbf 1\{T\le \mathcal C\}$. From these flags an analyst
constructs $K$ \emph{candidate indices}. In the general setting a candidate
differs in which conditions it includes, how they are coded or grouped, and
how they are weighted; in our simulation a candidate is a Cox model fit to
a fixed $d$-flag subset, so the $K$ candidates share a functional form and
differ only in their variable sets. Each candidate $m$ yields a fitted risk
score and an \emph{apparent} discrimination $\Chat_m$, evaluated on the
same data used to fit it. The analyst keeps the winner,
$\shat=\argmax_{1\le m\le K}\Chat_m$, the \emph{best-of-$K$} developed
index. This selection step is the source of the problem: the maximum of $K$
noisy quantities drifts upward, so $\Chat_{\shat}$ overstates the winner's
true quality, and the inflation grows with $K$.

\subsection{The fixed comparator}\label{setup-comparator}

The developed index is benchmarked against a \emph{fixed comparator}: a
linear score with frozen weights, applied as-is and
\emph{never re-fit} on the development sample; write $C_{\mathrm{comp}}$ for its concordance. This is our stand-in for an
off-the-shelf Charlson or Elixhauser index. The two arms are asymmetric by
design---the developed arm is both \emph{selected and fit} on the data
(hence optimistic), while the comparator carries no optimism. Valid
inference on their difference must respect this asymmetry as well as the
selection over $K$.

\subsection{Estimand: a post-selection difference in
concordance}\label{setup-estimand}

The target is the incremental discrimination of the index actually chosen,
$\Delta = C_{\shat}-C_{\mathrm{comp}}$, where $C_{\shat}$ is the
\emph{true} (population) concordance of the fitted winner. This is a
\emph{post-selection / conditional} (PoSI) quantity \citep{Berk2013}: the
truth \emph{for the candidate that won this replication}, not a
selection-averaged summary. The marginal alternative
$\bar\Delta=\mathbb E[C_{\shat}]-C_{\mathrm{comp}}$ is a distinct object;
the two coincide only when the candidates are exchangeable in true value
(the homogeneous regime) and diverge otherwise. This distinction is not
cosmetic: the bootstrap methods anchor to the conditional $\Delta$, whereas
the nested cross-validation comparators (\texttt{cv}, \texttt{cvclt},
\texttt{bht}) pool out-of-fold predictions across folds in
which different candidates may win and therefore estimate the marginal
$\bar\Delta$. A head-to-head coverage table that scored these targets
against one another without saying so would not be interpretable, so we
state the estimand up front.

\subsection{Handling censoring}\label{setup-censoring}

Harrell's $\Chat$ under right censoring is \emph{not} consistent for the
uncensored population concordance $C_m$; it converges to a
censoring-distribution-dependent functional $C_m^{H,G}$
\citep{Uno2011, GonenHeller2005}. The gap $\Xi_m=C_m^{H,G}-C_m$ is
\emph{score-specific}---it depends on how a given model orders subjects
relative to their censoring times---so the winner and the comparator, which
order subjects differently, incur \emph{different} shifts. In general $\Xi$
does \emph{not} cancel in $\Delta$: a common additive shift determined only
by the censoring distribution would be the exception, not the rule. Two
consequences follow. First, the \emph{confirmatory analysis uses Uno's IPCW
$C$}, a censoring-free target, at the cost of an estimated-weight nuisance
requiring a correctly specified censoring model; a censoring-model
sensitivity analysis confirms the direction and magnitude of the shift.
Second, whichever concordance functional is adopted, the truth set must be
scored consistently with the estimator. Crucially, the selection mechanism
this paper characterizes is \emph{orthogonal} to this choice---the winner's
curse operates on whichever concordance functional is used---so the
structural results carry over; only the estimand and its truth definition
must be kept internally consistent.

\section{Methods Compared}\label{methods}

\emph{Aim.} This section defines the six interval procedures we compare and the single
fitted quantity they all target, so that the theory of Section~\ref{theory} and the
results of Section~\ref{simulation} can be read against a common template. Of the six,
the \emph{selection-aware} bootstrap is the proposed fix; the others---a naive floor,
current single-step practice, and three cross-validation intervals spanning
anti-conservative to over-conservative---serve as reference points. Table~\ref{tab:methods}
is the at-a-glance summary.

We contrast six procedures for constructing a 95\% confidence interval
(CI) for $\Delta$. All six are applied to the \emph{same} fitted
quantity---the apparent difference
$\Dhat = \Chat_{\shat} - \Chat_{\mathrm{comp}}$, where
$\shat = \argmax_m \Chat_m$ is the candidate with the highest apparent
concordance on the development sample and the comparator is a fixed linear
score that is never re-fit. They differ only in how they resample and in
what optimism they attempt to remove. Throughout, ``replay the selection''
means recomputing $\argmax_m$ \emph{inside} a resample; ``hold the
comparator fixed'' means the frozen off-the-shelf score is re-evaluated but
never re-estimated.

\subsection{The three bootstrap procedures}\label{methods-boot}

All three bootstrap methods draw $B$ resamples of the development data and,
for the two corrected variants, form a \emph{location-shifted percentile
interval} \citep{Noma2021}: the percentile CI of the apparent $\Dhat$ is
translated by an estimated optimism $\widehat{\mathrm{Opt}}$, giving
$[\Dhat - \widehat{\mathrm{Opt}} - q_{.975},\ \Dhat -
\widehat{\mathrm{Opt}} - q_{.025}]$, where $q_\cdot$ are percentiles of the
centered bootstrap difference. The methods diverge in whether the selection
step is inside or outside the resampling loop.

\begin{itemize}
\item \textbf{naive} --- a percentile bootstrap of the apparent $\Dhat$
that \emph{re-selects} best-of-$K$ in each resample but applies \emph{no}
optimism correction. It is an uncorrected floor, expected to fail; it
isolates how much of the miscoverage is optimism rather than sampling
variability.
\item \textbf{single\_step} --- the standard in-house practice. It estimates
the Efron--Gong bootstrap optimism \citep{EfronGong1983} of the
\emph{already-selected} model $\shat$---Noma's location-shifted
optimism-corrected interval applied to the fixed winner---treating it
\emph{as if it were the only model ever fit}: the fixed winner is refit and
validated across bootstraps, but the $\argmax$-over-$K$ step is \emph{not}
replayed. By Proposition~1 this recovers the within-model
optimism $b_n$ but omits the selection term $S(K)$ entirely, which is the
source of its under-coverage.
\item \textbf{selection\_aware} \emph{(proposed)} --- identical machinery,
except best-of-$K$ is \emph{re-selected inside every bootstrap} (comparator
held fixed). Replaying the argmax makes the optimism estimate include the
expected maximum, so it targets $b_n + S(K)$ and corrects
$\mathrm{Opt}(K)$, at no additional compute over \texttt{single\_step} (the two
share one resampling loop). The correction is valid under exchangeability (where the
conditional and marginal targets coincide); at exact ties the
$n$-out-of-$n$ bootstrap of an argmax is delicate
\citep{Andrews2000, FangSantos2019}, and an $m$-out-of-$n$ / subsampling
variant is the principled route \citep{PolitisRomanoWolf1999, BickelSakov2008}
(Section~\ref{theory}).
\end{itemize}

\subsection{Nested cross-validation: a validity bracket}\label{methods-cv}

The remaining three procedures are \emph{nested}-CV comparators: each
re-selects best-of-$K$ inside every fold and turns the cross-validated
$C$-difference into an interval, differing only in \emph{which} variance
it uses. Rather than pit the proposed bootstrap against a single
cross-validation competitor, we let the three \emph{bracket} the validity
spectrum---anti-conservative (\texttt{cv}), calibrated (\texttt{cvclt}),
over-conservative (\texttt{bht})---so that matching the calibrated interior
corroborates the bootstrap's coverage rather than against a calibrated reference.

\textbf{cv} (repeated $V$-fold CV $+$ DeLong difference) forms the developed
score's concordance from pooled out-of-fold risks and pairs it against the
fixed comparator, using Uno's IPCW $C$ where the confirmatory estimand
applies. Its interval is the \emph{DeLong-type variance of the difference of
two $C$-statistics} \citep{DeLong1988}, the influence-function covariance of
the concordance estimator. Because this variance \emph{conditions on the
estimated out-of-fold risk score and omits the CV/selection-variability
term}, \texttt{cv} is \emph{anti-conservative by construction}, and it occupies the permissive end of the
bracket. (The DeLong difference variance can further degenerate in \emph{nested}
comparisons \citep{Demler2012}; the present developed-versus-frozen comparison is not
nested, so the anti-conservativeness rests on the omitted selection-variability term.)

\textbf{cvclt} keeps the same fold-level $C$-differences but replaces the
DeLong variance with a \emph{fold-level cross-validation standard error}: the
across-fold standard deviation of the per-fold $C$-difference over $\sqrt{V}$,
averaged over repeats. \citet{NadeauBengio2003} study exactly this across-fold
variance and show its uncorrected form is anti-conservative under train-set
overlap; we therefore do not rely on it being valid \emph{a priori} but
certify it empirically, from its coverage (below). Restoring the across-fold variability that
\texttt{cv} discards, it is empirically \emph{calibrated}---near
nominal and stable in $K$---so it is the \emph{valid} interior of the
bracket and the reference against which the proposed bootstrap is
judged.\footnote{Despite the name, \texttt{cvclt} is \emph{not} the cross-validation central
limit theorem of \citet{Bayle2020, AusternZhou2020}, whose per-\emph{observation} loss
variance is undefined for a concordance (a degree-2 $U$-statistic); its validity is certified
empirically here, from its coverage, not by appeal to that theorem.} Its point estimate averages the
$C$-difference \emph{within} folds rather than pooling all pairs across
folds, so it differs slightly from \texttt{cv}'s point---the two are not one
estimator with only the variance changed.

\textbf{bht} (a delete-a-block jackknife) partitions the development sample
into blocks, deletes each block in turn, and forms the interval from the
jackknife dispersion of the resulting cross-validated $C$-differences. It is
valid but \emph{over-conservative}, occupying the cautious end of the
bracket.

All three nested-CV intervals pool or average across folds in which
\emph{different} candidates may win, so all three target the
selection-\emph{averaged} $\bar\Delta$, not the conditional $\Delta$ of the
one index actually developed. The two coincide under exchangeability
(the homogeneity parameter $\eta{=}1$; $\eta\in[0,1]$ tunes how much true value the
candidates share, defined in \S\ref{sim-dgp}) and diverge in principle otherwise, though in our design
the divergence is too small to separate coverage (Web Appendix~C); the
proposed bootstrap anchors to the conditional $\Delta$ at $\eta{=}1$; in our design the two
targets do not diverge enough to separate coverage. A full nested-CV variance
\citep{BatesHastieTibshirani2024}, and its Cox-model test-error form
\citep{SunTibshirani2023}, is a further refinement not implemented here.

\medskip
\noindent Collecting the two families, Table~\ref{tab:methods} places all six procedures
side by side (an overview, not a results table).

\begin{table}[htbp]\centering
\begin{threeparttable}
\caption{The six interval procedures compared. All are applied to the same apparent
difference between the selected developed index and the fixed comparator, and differ in
whether the best-of-several selection is repeated (``replayed'') inside the resampling
loop and in what variance is used. The three bootstrap methods target the conditional
post-selection added value $\Delta$; the three cross-validation methods
(\texttt{cv}, \texttt{cvclt}, \texttt{bht}) target the marginal added value $\bar\Delta$
and \emph{bracket} the validity spectrum, from anti-conservative through calibrated to
over-conservative.}
\label{tab:methods}
\footnotesize
\setlength{\tabcolsep}{4pt}
\begin{tabular}{@{}llcclcll@{}}
\toprule
Method & Resamples & Best-of-$K$\tnote{a} & Optimism\tnote{b} & Interval\tnote{c} & Cost\tnote{e} & Target\tnote{d} & Role \\
\midrule
naive & bootstrap & replayed & none & percentile & $1\times$ & $\Delta$ & failing floor \\
single-step & bootstrap & winner fixed & $b_n$ & location-shift & $1\times$ & $\Delta$ & standard practice \\
selection-aware & bootstrap & \textbf{replayed} & $b_n + S(K)$ & location-shift & $1\times$ & $\Delta$ & proposed fix \\
cv & repeated $V$-fold & per split & out-of-fold & DeLong var. & $0.2\times$ & $\bar\Delta$ & anti-conservative \\
cvclt & repeated $V$-fold & per split & out-of-fold & fold-level CV SE & $0.2\times$ & $\bar\Delta$ & calibrated (valid) \\
bht & block jackknife & per split & out-of-fold & block jackknife & $2.3\times$ & $\bar\Delta$ & over-conservative \\
\bottomrule
\end{tabular}
\begin{tablenotes}[flushleft]\footnotesize
\item[a] How the best-of-$K$ choice is handled inside resampling. \emph{Replayed}: the
best-of-$K$ index is re-selected afresh in every resample, so selection variability
enters the interval. \emph{Winner fixed}: the index chosen on the full sample is held
fixed across resamples. \emph{Per split}: each cross-validation split re-selects its own
best-of-$K$.
\item[b] Optimism correction subtracted from the apparent difference. $b_n$ is the
ordinary single-model optimism (in-sample minus out-of-sample) for a fixed model;
$S(K)$ is the additional winner's-curse term from taking the best of $K$ candidates;
\emph{out-of-fold} removes optimism by evaluating on held-out folds.
\item[c] How the interval is formed. \emph{Percentile}: bootstrap percentile interval.
\emph{Location-shift}: re-centre by the estimated optimism, then use the bootstrap
spread. The remaining entries name the variance estimate used.
\item[d] $\Delta$ is the conditional post-selection added value of the single index
actually developed and reported; $\bar\Delta$ is the marginal added value averaged over
the whole selection procedure.
\item[e] Relative single-core wall-time, measured on the NHANES 2013--2014 cohort
($n{=}300$ development draw, $K{=}20$, $B{=}500$; mean of 3 draws). The three bootstrap
intervals are produced by one shared resampling loop, so the $1\times$ is the cost of all
three together; the block jackknife re-runs the whole cross-validation $1{+}G$ times. The
multiples depend on $B$ (a larger bootstrap raises the shared baseline).
\end{tablenotes}
\end{threeparttable}
\end{table}

\section{Theory: Why Single-Step Under-Covers, and Why Re-Selection Repairs
It}\label{theory}

This section states the mechanism as four propositions. The engine is a
single object---the joint asymptotic normality of the vector of candidate
concordances, so that the apparent winner is the \emph{maximum of
correlated Gaussians}. Everything follows from reading the winner's-curse
optimism as an expected maximum whose size, in standard-error units, does
not shrink with the sample. We give the intuition and a proof sketch for
each proposition and carry a per-result rigor rating
(proved/measured/open); full statements, regularity conditions, and proofs
are in Web Appendix~A. \emph{An applied reader may take the one conclusion that
matters---the miscalibration does not shrink as the cohort grows, so more data will not fix
it---and skip to Section~\ref{simulation}.}

\textbf{The engine (Assumption A1).} In words: each candidate's concordance is approximately
normal around its true value, and the $K$ candidates are jointly normal because they share
comorbidity flags---so the apparent winner is the maximum of correlated normal draws.
Formally, for a \emph{fixed} candidate set, the
$(K{+}1)$-vector $\sqrt n\big(\Chat_1 - c_1,\dots,\Chat_K -
c_K,\Chat_{\mathrm{comp}} - c_{\mathrm{comp}}\big) \Rightarrow N(0,\Sigma)$,
each $\Chat_m$ being an asymptotically linear functional of degree-2
$U$-statistics (Hájek projection plus delta method;
\citealp{DeLong1988, Uno2011, Kang2015}). This holds under a positive
comparable-pair probability, a unique pseudo-true Cox parameter with
positive-definite information \citep{StruthersKalbfleisch1986, LinWei1989},
non-degeneracy ($C_m \neq \tfrac12$), and distinct subsets so $\Sigma$ is
positive definite. Consequently the apparent winner
$\Chat_{\shat} = \max_m \Chat_m$ is the maximum of a Gaussian
vector---extreme-value territory. This step is proved under the stated
conditions, which our design-generating process meets.

\textbf{Proposition 1 (optimism decomposition).}\label{prop:p1} The
optimism of the apparent winner splits into a within-model term and a
selection term,
\begin{equation}
\mathrm{Opt}(K) \;=\; \mathbb{E}\!\left[\Chat_{\shat} - C_{\shat}\right] \;=\; b_n \;+\; S(K),
\label{eq:decomp}
\end{equation}
where $b_n = \mathrm{Opt}(1)$ is the ordinary single-model optimism and
$S(K)$ is the \emph{selection optimism}. Single-step estimates $b_n$ and
omits $S(K)$ entirely; $b_n$ is a common location shift that, \emph{under
exchangeability} (Web Appendix~A), cancels in $S(K) = \mathrm{Opt}(K) -
\mathrm{Opt}(1)$. The decomposition is proved given A1, and the $b_n$
characterization holds under the conditions of the binary design.

\textbf{Proposition 2 (the selection term is an expected maximum).} In the
homogeneous limit,
\begin{equation}
S(K) \;=\; \frac{\sigma}{\sqrt n}\,\sqrt{1-\rho_{\mathrm{eff}}(K)}\;a_K,
\qquad a_K = \mathbb{E}\big[\max_{1\le m\le K} Z_m\big],\ Z\sim N(0,I_K),
\label{eq:sk}
\end{equation}
with $a_K$ the finite-$K$ expected maximum ($\sqrt{2\log K}$ is only its
asymptote and overstates it for $K\le100$). For exact equicorrelation
$\rho$ the factor $\sqrt{1-\rho}\,a_K$ is exact; for general $\Sigma$
arising from overlapping flag subsets we \emph{approximate} the dependence by
an effective correlation $\rho_{\mathrm{eff}}(K)$ using standard Gaussian-comparison
inequalities (Slepian / Sudakov--Fernique) that bound the expected maximum of a correlated
Gaussian vector by that of an equicorrelated reference---a bound-based approximation, not an
identity. As $K$ grows the top competitors driving the maximum are increasingly correlated,
so $\rho_{\mathrm{eff}}$ tends to rise and $S(K)$ decelerates relative to the
i.i.d.\ rate but still \emph{grows with} $K$. The equicorrelation form is
exact; the general-$\Sigma$ reduction is leading-order; the constants are
measured.

\textbf{Proposition 3 (the coverage gap is $O(1)$ and non-vanishing in $n$ --- the
central result).}\label{prop:p3} Single-step's interval is centered at
$\Dhat - \hat b_n$---miscentered from $\Delta$ by exactly the omitted
$S(K)$---with half-width $z_{.975}\,\tau_n$, where $\tau_n$ is the standard
error of the difference. Both $S(K)$ and $\tau_n$ are $O(1/\sqrt n)$ CLT
quantities, so their ratio is scale-free:
\begin{equation}
\dstar \;=\; \frac{S(K)}{\tau_n} \;=\; O(1), \quad n\text{-invariant, growing like } a_K.
\label{eq:dstar}
\end{equation}
The consequence is that the under-coverage of single-step is
\emph{structural, not a small-sample artifact}---it does not disappear as
$n\to\infty$; it is a fixed, $K$-driven miscalibration. \emph{The constant,
derived.} Writing $\gamma = \sigma_{\mathrm{comp}}/\sigma$ and $\rho'$ for
the candidate--comparator correlation,
$\tau_n = (\sigma/\sqrt n)\sqrt{1+\gamma^2 - 2\gamma\rho'}$, giving the
closed form
$\dstar(K) = a_K\sqrt{1-\rho}/\sqrt{1+\gamma^2 - 2\gamma\rho'}$, which
equals $a_K/\sqrt2$ \emph{exactly} under variance-homogeneity
($\gamma=1,\rho'=\rho$)---an idealization the measured $\gamma\approx0.85$,
$\rho'\approx0.40$ approach but do not meet. Plugged into the closed form (with
the mean-pairwise $\rho\approx0.5$; the larger effective $\rho_{\mathrm{eff}}\approx0.6$ argued to govern
the maximum would lower the prediction, so the point match is order-of-magnitude), these
give a predicted $\dstar(20)$ near
the observed $\approx1.37$ (Table~\ref{tab:dstar}), consistent to within the $\sim$10--20\% sampling wobble of
the constants---a plug-in from the independently measured constants (via the
equicorrelation approximation), not a tuned fit; the robust,
constant-free content is that $\dstar$ is $O(1)$ and does not shrink with $n$.
That $\dstar=O(1)$ and non-vanishing in $n$ (bounded away from zero, not shrinking toward nominal) is proved under conditions and confirmed
by measurement; the point constant is measured to within the
covariance-constant wobble.

\textbf{Boundary corollary (Type-I).} At the least-favorable null
$\Delta = 0$, single-step's one-sided false-superiority rate is
$\approx \bar\Phi(z_{.975} - \dstar_{\text{1-sided}}) \gg \alpha$ (with $\bar\Phi=1-\Phi$) and rises
with $K$: measured $0.30$ at $K{=}100$ against a nominal $0.025$. This is
the mechanism by which a truly non-superior index is declared superior.

\textbf{Proposition 4 (the fix, and the exact-ties caveat).} The
\emph{selection-aware} interval replays the $\argmax$ inside every
bootstrap, so its optimism estimate includes $\mathbb{E}[\max]$ and targets
$b_n + S(K)$, correcting $\mathrm{Opt}(K)$. \emph{Caveat.} A
reselection bootstrap estimates the optimism of the \emph{selection
algorithm} (a marginal quantity), which equals the correction for the
\emph{conditional} winner only under exchangeability; off homogeneity the
two differ. At exact ties the $n$-out-of-$n$ bootstrap of an $\argmax$ is
delicate \citep{Andrews2000, FangSantos2019}; whether this bites for a
bootstrapped \emph{location shift}---a smooth functional of
$\mathbb{E}[\max]$, not the max's sampling law---is the crux; empirically
the method is well-calibrated in the exchangeable regime, consistent with
location-shift robustness, with an $m$-out-of-$n$ / subsampling variant
recommended at exact ties. This is proved under conditions away from exact
ties and scoped at exact ties.

One question remains open: a full
consistency proof for selection-aware at \emph{exact} ties (location-shift versus
argmax-law, with the explicit subsampling rate $m\to\infty$, $m/n\to0$, or
the conditional-inference-on-winners route of Andrews--Kitagawa--McCloskey
\citep{AKM2024}). The two once-open companions are now in hand: the $\dstar$
constant has the derived closed form above, and the normal approximation to the maximum is
accurate at the usual $O(n^{-1/2})$ rate (a fixed-$K$ Berry--Esseen bound for the maximum of
$K$ averages; Web Appendix~A).

\section{Simulation Study}\label{simulation}

We validate the four main-grid interval methods (naive, single-step,
selection-aware, \texttt{cv}) against a fully known truth, in a
design built so that the sample optimism of a best-of-$K$ index, the
selection component $S(K)$, and the target difference $\Delta$ are all
computable exactly. The full data-generating process is in Web
Appendix~B; the complete grid is in Web
Appendix~C.

\subsection{Known-truth data-generating process}\label{sim-dgp}

Each development sample of size $n$ carries $p=20$ binary comorbidity flags
with marginal prevalence $0.25$ and an exchangeable latent-Gaussian
correlation of $0.20$. Survival times are exponential with hazard
$\propto \exp(X^\top\beta)$, where exactly $6$ of the $20$ flags are true
signal at $\log 1.6$ and the remaining $14$ are noise; censoring is
independent exponential, its rate calibrated per cell to a target
proportion. A \emph{candidate} is a Cox model on a fixed $d=8$-flag subset.
The candidate library is \emph{nested across $K$}: one ordered pool of
subsets is drawn once per base cell---with a seed independent of $K$---and
cell $K$ uses its first $K$ entries, so the $K$-trend is read cleanly
rather than confounded by re-randomised libraries. A homogeneity parameter $\eta\in[0,1]$ governs how much true value the candidates share: with probability
$\eta$ a candidate receives equal true value (pure winner's curse,
the argmax is noise-driven), otherwise a random count (heterogeneous true
value, so the winner partly \emph{earns} its edge). $\eta=1$ is
the least-favourable, worst-case cell for single-step. The \emph{comparator}
is a fixed linear score, never re-fit. Under the alternative it is a weak
oracle giving a genuine $\Delta>0$ (power regime); under the null it is
a matched-structure index estimated once on an independent same-size cohort
and then frozen, yielding a true $\Delta\approx 0$ at $\eta=1$---the
genuine least-favourable null against which a superiority test's size must
be judged. The \emph{truth} is the difference in concordance for the model
\emph{actually selected each replication} (a post-selection, conditional
target), computed on a large uncensored test set.

\subsection{Design sweep and reporting}\label{sim-sweep}

The full grid crosses $K\in\{1,5,20,100\}$, two signal conditions (alternative and null), $n\in\{100,150,300\}$, censoring fixed at $0.3$, and homogeneity $\eta\in\{1,0.5,0\}$, giving \textbf{72 cells}, at $R=1{,}000$
replications per cell with $B=500$ bootstraps each and $V{=}10$-fold CV
repeated $10$ times, run as a $720$-task array on UBC ARC. Replicate seeds
use common random numbers across $K$ \emph{and} across methods, so both the
cross-$K$ contrast and the method-vs-method contrasts are paired. For every
method in every cell we report coverage of the 95\% CI with its
Monte-Carlo SE ($\le 0.016$, about $0.007$ near nominal coverage, at $R{=}1{,}000$), the
\emph{declare-superior rate} ($=$ power under the alternative, Type-I under
the null), and the optimism decomposition; bias, width, and the
interval score are in Web Appendix~C. The validity-bracket comparison of
Table~\ref{tab:frontier}---which adds the \texttt{cvclt} and \texttt{bht}
intervals---uses a companion 24-cell run at the same design ($n=150$,
homogeneous and heterogeneous candidates) with $B=200$ bootstraps; the two runs
share the data-generating process and differ only in grid and bootstrap depth,
so single-step coverage differs only trivially between them (for example,
$0.94$ vs $0.92$ at $K=1$).

\subsection{Results: confirmation of the theory}\label{sim-results}

The simulation grid confirms each proposition. All numbers below are
$\eta=1$, $n=150$ unless noted; MC-SE $\le 0.016$ ($\approx 0.007$ near 0.95).

\begin{figure}[htbp]
\centering
\includegraphics[width=0.98\linewidth]{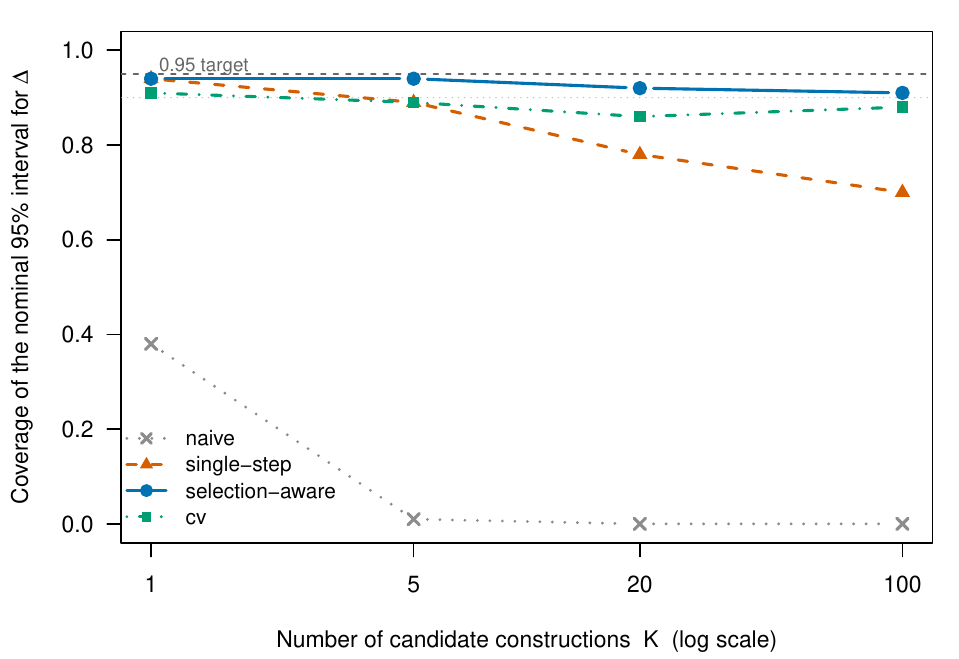}
\caption{Empirical coverage of the nominal 95\% interval for $\Delta$ against the number of candidates $K$ ($\eta=1$, $n=150$; 1{,}000 replications). The naive bootstrap collapses once selection is present; single-step (standard practice) falls monotonically from near-nominal at $K=1$ to about 0.70 at $K=100$; the proposed selection-aware bootstrap holds near nominal across $K$; the repeated-cross-validation benchmark is mildly and non-monotonically anti-conservative (coverage 0.86--0.91 across $K$), below selection-aware throughout and above single-step once selection bites ($K\ge20$). The dashed line marks the 0.95 target.}
\label{fig:coverage}
\end{figure}

\textbf{P1/P3 --- single-step under-covers, worsening monotonically in $K$;
the fix holds} (Table~\ref{tab:coverage}, Figure~\ref{fig:coverage}). The naive percentile bootstrap
collapses once selection is present; single-step, fine at $K{=}1$, falls to
$0.70$ at $K{=}100$; selection-aware holds near nominal throughout. On
coverage alone, selection-aware merely matches a valid cross-validation
interval---this parity is not itself the contribution. The residual optimism single-step misses
grows $0.006\to0.052$ across $K{=}1\to100$, tracking the growing winner's
curse, whereas selection-aware's residual stays $0.006\to0.020$: it removes
the piece single-step ignores.

\begin{table}[htbp]\centering
\caption{Coverage of the nominal 95\% confidence interval for $\Delta$, by number of candidates $K$ ($\eta=1$, $n=150$; target 0.95; MC-SE $\le 0.016$, $\approx 0.007$ near 0.95). Single-step collapses in $K$ while selection-aware holds near nominal.}
\label{tab:coverage}
\begin{tabular}{@{}rrrrr@{}}
\toprule
$K$ & naive & single-step & selection-aware & cv \\
\midrule
1   & 0.38 & 0.94 & 0.94 & 0.91 \\
5   & 0.01 & 0.89 & 0.94 & 0.89 \\
20  & 0.00 & 0.78 & 0.92 & 0.86 \\
100 & 0.00 & 0.70 & 0.91 & 0.88 \\
\bottomrule
\end{tabular}

\end{table}

\textbf{P3 (structural) --- the bias-to-SE ratio $\dstar$ is $O(1)$ and does
not vanish with $n$} (Table~\ref{tab:dstar}). $\dstar$ grows with $K$ (like the
expected maximum $a_K$) yet stays $O(1)$ across $n$---with a mild
upward drift (partly a two-run splice), never toward zero; single-step coverage likewise does
not recover ($0.78,0.78,0.79$ at $K{=}20$). This non-shrinkage is the
signature that separates \emph{structural} under-coverage from a small-sample
artifact. A dedicated large-$n$ run extends it over a twentyfold sweep: at
$K{=}20$, $\dstar$ holds at $1.5$--$1.7$ from $n{=}600$ to $2{,}000$, and at
$K{=}100$ at $1.8$--$2.0$, over which single-step coverage never
recovers---ruling out a small-sample explanation. (The tabulated $\dstar$
carries a small non-vanishing floor visible at $K{=}1$, a residual
optimism-estimation term the selection theory $S(1){=}0$ does not model. The
selection-only component $\dstar(K){-}\dstar(1)$ [Table~\ref{tab:dstar}, last
rows] nets out this floor and is roughly flat within each run, so the
structural signature is the \emph{persistence} of the selection ratio, not
exact $n$-invariance of the raw one.)

\begin{table}[htbp]\centering
\caption{The bias-to-standard-error ratio $\dstar=$ residual-bias/SE (the alternative), by $K$ and sample size $n$. $\dstar$ grows with $K$ but stays $O(1)$ across a \emph{twentyfold} range of $n$ ($100\to2{,}000$). The right block ($n\ge600$) is a dedicated large-$n$ run with $K\in\{1,20,100\}$. The last two rows give the selection-only component $\dstar(K){-}\dstar(1)$, which nets out the $K{=}1$ optimism-estimation floor and is roughly flat within each run.}
\label{tab:dstar}
\begin{tabular}{@{}rrrrrrr@{}}
\toprule
$K$ & $n{=}100$ & $n{=}150$ & $n{=}300$ & $n{=}600$ & $n{=}1000$ & $n{=}2000$ \\
\midrule
1   & 0.23 & 0.20 & 0.13 & 0.12 & 0.26 & 0.37 \\
5   & 1.03 & 0.97 & 0.81 & ---  & ---  & ---  \\
20  & 1.37 & 1.37 & 1.29 & 1.54 & 1.62 & 1.66 \\
100 & 1.64 & 1.63 & 1.53 & 1.80 & 1.85 & 1.96 \\
\addlinespace
\multicolumn{7}{@{}l}{\footnotesize\emph{selection-only $\dstar(K){-}\dstar(1)$ (nets out the $K{=}1$ optimism floor):}}\\
$20{-}1$  & 1.14 & 1.17 & 1.16 & 1.42 & 1.36 & 1.29 \\
$100{-}1$ & 1.41 & 1.43 & 1.40 & 1.68 & 1.59 & 1.59 \\
\bottomrule
\end{tabular}

\end{table}

\textbf{Validity check and estimand
distinction} (Table~\ref{tab:frontier}). The reframed comparison places
selection-aware inside a \emph{validity bracket} of three nested-CV
competitors, each of which re-selects best-of-$K$ in every fold:
\texttt{cv} (repeated-CV $+$ DeLong difference variance), anti-conservative
by construction; \texttt{cvclt}, a calibrated fold-level
cross-validation standard error (a naive across-fold variance in the style
of \citet{NadeauBengio2003}, whose validity we certify \emph{empirically}
from its coverage---not by appeal to the observation-level CV-CLT of
\citet{Bayle2020} and \citet{AusternZhou2020}, which is undefined for a
concordance, a degree-2 $U$-statistic); and \texttt{bht}, a delete-a-block
jackknife of the CV $C$-difference. At $\eta=1$, $n{=}150$ these
bracket the validity spectrum---coverage runs \texttt{cv} $\approx0.89$
(under) $<$ selection-aware$/$\texttt{cvclt} $\approx0.92$--$0.93$
(near-nominal) $<$ \texttt{bht} $\approx0.98$ (over)---while single-step
collapses $0.92\to0.68$ across $K{=}1\to100$. Two facts do the work.
\emph{(i) Validity by matching.} Selection-aware's
coverage ($0.92,0.92,0.92,0.91$ across $K{=}1,5,20,100$) matches
\texttt{cvclt}'s ($0.93,0.92,0.93,0.93$), a competitor separately certified
as valid, so its near-nominal, $K$-stable coverage is genuine calibration
rather than a win over an anti-conservative benchmark; it is near-nominal
and stable in $K$, \emph{not} exactly $0.95$.
Re-scoring every method at a \emph{common} Type-I from the saved
per-replication records, selection-aware is at least as powerful as
\texttt{cvclt} on the matched frontier: at level $0.05$ the paired power gap
(selection-aware minus \texttt{cvclt}) is $+0.01,+0.04,+0.02,+0.06$ across
$K{=}1,5,20,100$ (paired SE $\approx0.01$)---a small edge, significant at
$K{=}5$ and $K{=}100$ and a statistical tie at $K{=}1,20$---while against the
over-conservative \texttt{bht} (matched power $0.15,0.13,0.10,0.12$) it wins
strictly once selection is active ($K{\ge}5$). Absolute power is low by design ($n{=}150$, true
$\Delta\approx0.03$), so the comparison is \emph{relative} at
matched Type-I; raw declare-superior rates are not comparable across methods
that cover differently, and we do not report them as a winner.
\emph{(ii) The estimand distinction.} Selection-aware targets the
\emph{conditional} post-selection $\Delta$---the added value of the one
index actually developed and published---whereas \texttt{cv}, \texttt{cvclt},
and \texttt{bht} target the \emph{marginal} $\bar\Delta$, the average added
value of the whole selection \emph{procedure}. This is a distinction of
\emph{construction}, not of coverage in our design: a re-aggregation of the
per-replication truth shows the realized winner's true value varies little
across replications (across-replication SD $\le 0.02$, far below the interval
half-width), so the conditional and marginal targets stay close and
\texttt{cvclt} covers the \emph{conditional} $\Delta$ as well as
selection-aware at every homogeneity level (Web Appendix~C). In our design the
distinction is therefore conceptual rather than a coverage difference, while single-step
collapses throughout. This check---valid exactly
where a calibrated CV interval is valid, aimed at the estimand a
developer actually reports, and $K$-stable where single-step
collapses---together with the $\argmax$-over-$K$ optimism decomposition, is
the paper's positive contribution.

\begin{table}[htbp]\centering
\caption{The validity bracket and matched-Type-I frontier at $n{=}150$, $\eta=1$ ($R{=}1{,}000$). Rows are the interval procedures within each $K$ block; columns give coverage of the nominal 95\% interval under the alternative and power re-scored to a matched Type-I of $0.05$. The three nested-CV competitors bracket the validity spectrum: \texttt{cv} (repeated-CV $+$ DeLong difference) is anti-conservative, \texttt{cvclt} (a calibrated fold-level cross-validation standard error) is valid, and \texttt{bht} (delete-a-block jackknife of the CV $C$-difference) is over-conservative. Selection-aware's coverage matches the valid \texttt{cvclt}, and its matched-Type-I power meets or exceeds \texttt{cvclt} (paired edge significant at $K{=}5,100$) while beating \texttt{bht} once selection is active. Single-step is shown for reference only (its raw power rides on an inflated Type-I). MC-SE $\approx0.008$ for coverage and $\approx0.013$--$0.016$ for the matched-power columns; contrasts paired via common random numbers.}
\label{tab:frontier}
\begin{tabular}{@{}c l cc@{}}
\toprule
$K$ & Method & Coverage (alt) & Power @ $0.05$ \\
\midrule
\multirow{5}{*}{$1$}   & single-step     & 0.92 & 0.16 \\
                       & cv              & 0.92 & 0.15 \\
                       & cvclt           & 0.93 & 0.14 \\
                       & selection-aware & 0.92 & 0.16 \\
                       & bht             & 0.96 & 0.15 \\
\addlinespace
\multirow{5}{*}{$5$}   & single-step     & 0.87 & 0.22 \\
                       & cv              & 0.90 & 0.19 \\
                       & cvclt           & 0.92 & 0.18 \\
                       & selection-aware & 0.92 & 0.22 \\
                       & bht             & 0.97 & 0.13 \\
\addlinespace
\multirow{5}{*}{$20$}  & single-step     & 0.79 & 0.22 \\
                       & cv              & 0.89 & 0.16 \\
                       & cvclt           & 0.93 & 0.18 \\
                       & selection-aware & 0.92 & 0.20 \\
                       & bht             & 0.97 & 0.10 \\
\addlinespace
\multirow{5}{*}{$100$} & single-step     & 0.68 & 0.23 \\
                       & cv              & 0.89 & 0.17 \\
                       & cvclt           & 0.93 & 0.16 \\
                       & selection-aware & 0.91 & 0.22 \\
                       & bht             & 0.98 & 0.12 \\
\bottomrule
\end{tabular}

\end{table}

\textbf{Attenuation off $\eta=1$.} The failure is worst under the
pure winner's curse but does not vanish: single-step coverage at $K{=}100$
is $0.70$ ($\eta=1$), $0.81$ ($0.5$), $0.77$ ($0$)---never
nominal, consistent with the winner partly earning its edge as candidates
become heterogeneous.

\textbf{Robustness to the concordance metric (Uno's IPCW $C$).} Re-running
the design under Uno's censoring-free IPCW $C$ reproduces the coverage and
boundary-error findings. Single-step coverage still falls with $K$ ($0.93\to0.68$ across
$K{=}1\to100$) while selection-aware holds ($\approx0.93$); at the null
boundary the single-step false-superiority rate still climbs
($0.06\to0.30$) while selection-aware controls it ($\approx0.07$); and
selection-aware's near-nominal, $K$-stable coverage under Uno's $C$ mirrors
the Harrell headline. Because the selection mechanism (P1--P3)
operates on whichever concordance functional is used, the finding is
\emph{not} an artifact of Harrell's censoring-dependent estimand (Web
Appendix~C).

\section{Real-Data Confirmation}\label{application}

This section confirms the finding directly on real NHANES data, reading
discrimination ($C$) only. The \emph{decision} consequence---how often the
choice of interval changes a ``does my index beat Charlson?'' verdict---is
quantified with exact truth by the simulation's boundary Type-I inflation
(single-step $0.30$ vs selection-aware $0.07$ at $K{=}100$,
Section~\ref{simulation}) and illustrated, modestly, on the NHANES 2013--2014
cycle in Web Appendix~D. What we confirm \emph{here}, directly, is the
mechanism itself.

It is not confined to synthetic covariance: we run a
\emph{semi-synthetic finite-population coverage experiment on real NHANES
comorbidity--mortality covariance}. We pool the 2005--2010 NHANES
cycles of adults aged $\ge 40$ with linked public-use mortality
\citep{NHANES2013, NCHS_LMF} and fix a disjoint partition into a reference
set (REF, $n{=}2{,}500$), a large held-out truth set (EVAL, $n{=}4{,}000$),
and a development pool (POOL, $n{=}4{,}310$). Each replication draws a
development subsample from POOL, builds the best-of-$K$ index from a library
of $d{=}4$-flag Cox candidates, and scores it against a frozen comparator
on the fixed EVAL truth; the four main-grid interval methods of
Section~\ref{methods} are applied exactly as in the simulation, under both
Harrell's and Uno's $C$. All \emph{selection}---the near-exchangeable
sub-library membership and the null comparator---is measured \emph{off} the
EVAL truth (candidates fit on POOL, scored on REF), so no quantity used to
define the estimand sees the coverage truth. The design, provenance, and a
full adversarial pre-registration are in Web
Appendix~D. This is a coverage experiment on real
covariance, \emph{not} a clinical claim about NHANES mortality; real
ICD-based Charlson/Elixhauser coding is routed to a companion claims-based study.

The experiment reads two theory-predicted regimes side by side. The
\emph{natural, heterogeneous full library} is the common case; the
\emph{constructed near-exchangeable sub-library} is the real-data analog of
the simulation's $\eta\approx1$ danger zone (candidates within a
narrow true-concordance band). Results (Table~\ref{tab:nhanes}; primary
readout is coverage of the finite-cohort held-out true $\Delta$, valid for
any realized $\Delta$):

\begin{itemize}
\item \textbf{The winner's curse is large on real comorbidity data.} The
naive percentile bootstrap collapses at $K{=}20$ (coverage $0.09$--$0.25$ at
small $n$, against $0.83$--$0.98$ at $K{=}1$)---the selection inflation is
real, not a synthetic artifact.
\item \textbf{Single-step removes most of it on the natural library.} On the
heterogeneous full library single-step coverage stays near nominal
($0.89$--$0.96$, rising with $n$): this is the map's \emph{safe region},
where candidates resolve as the events-per-variable grows.
\item \textbf{$\dstar$ attenuates on the full library and persists on the
near library.} On the full library $\dstar$ attenuates roughly $1/\sqrt n$
($K{=}20$: $1.07\to0.37$ as $n{:}150\to1{,}200$; Uno $1.10\to0.57$), the
heterogeneity-attenuation signature as candidates resolve; on the near-exchangeable
library it \emph{persists at $O(1)$ and does not shrink with $n$}
($1.24, 1.34, 1.23, 1.21$ across the eightfold $n$ range; Uno
$1.27, 1.36, 1.25, 1.23$), reproducing the P3 mechanism on real data. The
attenuation is a \emph{change of regime} as $n$ grows---not $n$-recovery
within a fixed regime---and the near curve holding $O(1)$ near the
homogeneous-limit reference ($a_{20}/\sqrt2\approx1.32$) is the control that
rules out a small-sample explanation. This is order-1 corroboration of the
mechanism, not a parameter-free numeric match.
\item \textbf{The correction earns its keep in the danger zone.} On the near
library single-step under-covers at $K{=}20$ and does \emph{not} recover with
$n$ (coverage $0.89, 0.82, 0.84, 0.84$ Harrell; $0.88, 0.82, 0.84, 0.81$
Uno), while selection-aware repairs to near nominal ($0.92$--$0.98$) under
both metrics---the real-data echo of the simulation's exchangeable cells.
\item \textbf{Uno $\approx$ Harrell throughout}, so the pattern is not a
metric artifact; the fixed 4{,}000-subject held-out truth set removes shrinking-truth
and dependence artifacts.
\end{itemize}

\begin{table}[htbp]\centering
\caption{NHANES semi-synthetic coverage experiment, at the headline $K{=}20$ least-favourable null, by candidate library (natural heterogeneous \emph{full} vs constructed \emph{near}-exchangeable), metric, and sample size $n$; truth-blind selection, 500 replications (Web Appendix~D). \emph{Panel A:} the bias-to-SE ratio $\dstar$ attenuates $\sim 1/\sqrt n$ on the full library as candidates resolve (the heterogeneity-attenuation signature) but persists at $O(1)$ and does not shrink with $n$ on the near-exchangeable library (the P3 mechanism). \emph{Panel B:} single-step coverage tracks $\dstar$---near-nominal on the full library, but under-covering on the near library and \emph{not} recovering with $n$. \emph{Panel C:} selection-aware repairs the near cells to 0.92--0.98 under both metrics.}
\label{tab:nhanes}
\begin{tabular}{@{}ll rrrr@{}}
\toprule
 & & \multicolumn{4}{c}{sample size $n$} \\
\cmidrule(l){3-6}
Metric & Library & 150 & 300 & 600 & 1200 \\
\midrule
\multicolumn{6}{@{}l}{\emph{Panel A: bias-to-SE ratio $\dstar$}}\\
\multirow{2}{*}{Harrell} & full (heterogeneous) & 1.07 & 0.84 & 0.43 & 0.37 \\
                         & near-exchangeable    & 1.24 & 1.34 & 1.23 & 1.21 \\
\multirow{2}{*}{Uno}     & full (heterogeneous) & 1.10 & 0.89 & 0.56 & 0.57 \\
                         & near-exchangeable    & 1.27 & 1.36 & 1.25 & 1.23 \\
\addlinespace
\multicolumn{6}{@{}l}{\emph{Panel B: single-step coverage of the finite-cohort true $\Delta$ (0.95 target)}}\\
\multirow{2}{*}{Harrell} & full (heterogeneous) & 0.89 & 0.92 & 0.96 & 0.96 \\
                         & near-exchangeable    & 0.89 & 0.82 & 0.84 & 0.84 \\
\multirow{2}{*}{Uno}     & full (heterogeneous) & 0.89 & 0.92 & 0.95 & 0.96 \\
                         & near-exchangeable    & 0.88 & 0.82 & 0.84 & 0.81 \\
\addlinespace
\multicolumn{6}{@{}l}{\emph{Panel C: selection-aware coverage of the finite-cohort true $\Delta$ (the repair; 0.95 target)}}\\
\multirow{2}{*}{Harrell} & full (heterogeneous) & 0.96 & 0.94 & 0.92 & 0.91 \\
                         & near-exchangeable    & 0.97 & 0.95 & 0.95 & 0.92 \\
\multirow{2}{*}{Uno}     & full (heterogeneous) & 0.96 & 0.92 & 0.93 & 0.93 \\
                         & near-exchangeable    & 0.98 & 0.93 & 0.94 & 0.93 \\
\bottomrule
\end{tabular}

\end{table}

The construction and evaluation machinery
(candidate assembly, integer point-scoring, stability selection, and the
accuracy/optimism suite, including the IPCW Brier score
\citep{Graf1999}) is demonstrated on fully public datasets in the
accompanying repository. In sum, the present paper establishes and validates
the inferential procedure by simulation, confirms its when-it-matters map on
real NHANES comorbidity--mortality covariance above, and illustrates its
decision consequence on the 2013--2014 cycle in Web Appendix~D. Applying the
full correction to administrative-claims Charlson/Elixhauser coding across
disease subcohorts \citep{Quan2005} is left to future work.

\section{Discussion}\label{discussion}

\textbf{What this means for practice}: When a comorbidity index is
developed by trying several constructions and the best-scoring one is
carried forward, the reported advantage over a fixed Charlson/Elixhauser
comparator is optimistic in a way that standard single-step optimism
correction does not remove. That correction treats the winner as if it were
the only model ever fit; it estimates the within-model optimism $b_n$ but
omits the selection term $S(K)$ (Proposition~1). The practical
consequence is concrete and, in our simulations, large: at $K=20$
homogeneous candidates and $n=150$, single-step coverage of the 95\%
interval for $\Delta$ falls to 0.78 and its false-superiority rate at a true
$\Delta=0$ reaches roughly a third of replications at $K=100$ (0.30
versus a nominal 0.025). A team following ordinary practice would declare a
non-superior index superior far more often than its stated error rate
implies.

Two features of the problem determine when it bites, and both are visible
to analysts before any inference is run. First, the winner's curse is worst
when candidates are \emph{similar in true quality} (the homogeneous
regime): there the winner is chosen almost entirely by luck, so all of the
selection optimism is spurious. When candidates genuinely differ, the
winner partly earns its selection and the failure attenuates---though in our
runs it never fully disappears. Second, the miscalibration grows with the
number of candidates $K$ and is worse at small events-per-variable, because
both enlarge the selection term relative to the sampling error. The
bias-to-standard-error ratio $\dstar = S(K)/\tau_n$ grows like the expected
maximum of $K$ correlated Gaussians and---the structural point---is
essentially bounded and non-vanishing in $n$. The under-coverage is therefore not a small-sample
artifact that a larger cohort will cure; it is a fixed, $K$-driven
miscentering (Proposition~3), and the NHANES experiment shows
the same regime map on real comorbidity covariance.
Table~\ref{tab:rot} distils this into a quick operating-characteristics guide
an analyst can consult before running the correction.

\begin{table}[htbp]\centering
\caption{Rule of thumb for the near-exchangeable worst case (candidates of
similar true quality, $n{=}150$): as the number of tried constructions $K$
grows, single-step's nominal $95\%$ interval covers less and its
false-superiority rate at a true null ($\Delta{=}0$) climbs, while
selection-aware holds both. Values from the validity-bracket run
($\eta=1$): the bracket-run counterpart of the main grid's
$0.94\to0.70$ single-step coverage (Table~\ref{tab:coverage}) and its
boundary false-superiority at $K{=}100$ ($0.33$ vs $0.09$ here; $0.30$ vs $0.07$ on the
main grid---the two runs agree within Monte-Carlo error). Reading: try more constructions of
similar quality on a small cohort, and expect this much hidden inflation.}
\label{tab:rot}
\begin{tabular}{@{}r cc cc@{}}
\toprule
 & \multicolumn{2}{c}{single-step} & \multicolumn{2}{c}{selection-aware} \\
\cmidrule(lr){2-3}\cmidrule(lr){4-5}
$K$ tried & 95\% cov. & false-sup. & 95\% cov. & false-sup. \\
\midrule
$1$   & 0.92 & 0.06 & 0.92 & 0.06 \\
$5$   & 0.87 & 0.15 & 0.92 & 0.06 \\
$20$  & 0.79 & 0.24 & 0.92 & 0.08 \\
$100$ & 0.68 & 0.33 & 0.91 & 0.09 \\
\bottomrule
\end{tabular}
\end{table}

Our recommendation is correspondingly simple: \textbf{when more than one
index construction was tried, report a selection-aware interval}---fold the
``pick the best of $K$'' step into the bootstrap while holding the
off-the-shelf comparator fixed. This adds one resampling loop---refitting the
$K$ candidates on each resample ($K\cdot B$ fits per analysis, seconds to
minutes for the small Cox models at issue)---and it restored coverage close to nominal across the grid
(0.90--0.94) and held the boundary false-superiority rate to 0.07---far
below single-step's 0.30, though still about three times the 0.025 one-sided
nominal. Where only a single, pre-specified index is compared ($K=1$),
single-step and selection-aware coincide and the correction is unnecessary.

\textbf{Positioning and scope}: We specify what selection-aware does and does not provide. Its comparative case rests on two legs. The first is
an \emph{empirical validity check paired with an estimand distinction},
established against a \emph{bracket} of nested cross-validation competitors
that span the validity spectrum. All three re-select the best of $K$ in
every fold---so they target the \emph{marginal} added value $\bar\Delta$ of
the selection procedure---and differ only in how fold-level variability
becomes an interval: \texttt{cv} (repeated CV $+$ DeLong difference
variance) is anti-conservative by construction ($\approx0.89$);
\texttt{cvclt}, a calibrated fold-level cross-validation standard
error in the sense of \citet{NadeauBengio2003}, is near-nominal and stable
in $K$ ($\approx0.92$--$0.93$); and \texttt{bht}, a delete-a-block
jackknife of the cross-validated $C$-difference, is valid but
over-conservative ($\approx0.97$--$0.98$, and about twice as wide as
\texttt{cvclt} at $K{=}100$: $0.245$ vs $0.133$). Read against this bracket,
the coverage result is an empirical validity check against calibrated competitors:
selection-aware's $\approx0.92$ coverage \emph{matches} \texttt{cvclt}, the
calibrated member, and holds there as $K$ grows while single-step collapses.
On the \emph{matched}-Type-I frontier---where the comparison must be
relative, since absolute power is low at $n=150$ with a true
$\Delta\approx0.03$---selection-aware is at least as powerful as
\texttt{cvclt} (a paired power advantage of $+0.01$ to $+0.06$ across $K$,
MC-SE $\approx0.01$, significant at $K{=}5$ and $K{=}100$ and statistically
equivalent otherwise) and, once selection is active ($K{\ge}5$), strictly
more powerful than the over-conservative \texttt{bht}. A plausible reason it can match
a \emph{valid} cross-validation interval's power is that each bootstrap
resample uses all $n$ observations while every cross-validated estimator
discards its held-out fold.

The bracket also answers the natural rejoinder to any selection
correction---``why not simply cross-validate?''---and the answer is about
\emph{estimand}, not calibration. \texttt{cvclt} shows that a valid CV
interval exists, but it targets the marginal $\bar\Delta$, the average added
value of the selection \emph{procedure}; selection-aware instead targets the
conditional post-selection $\Delta$, the added value of the one index an
investigator actually developed, published, and will deploy. When candidates
are exchangeable ($\eta=1$) the two estimands coincide---the clean
head-to-head regime---so the coverage and frontier comparisons there are
genuine like-for-like. Off exchangeability the two estimands diverge in
principle, but in our design that divergence is small (the realized winner's
true value varies little across replications), so the calibrated
cross-validation interval also covers the conditional truth near-nominally;
the distinction is therefore one of \emph{construction}---selection-aware
targets the conditional $\Delta$ directly at $\eta{=}1$---a conceptual distinction rather
than a coverage gap in these cells.

The second leg---the \emph{argmax-over-$K$ optimism decomposition}---names
and quantifies a failure mode that final-model-only correction silently omits.
This is the cautionary contribution, and it is the reason the
recommendation is actionable rather than merely diagnostic.

\textbf{Limitations and open theory}: Several caveats are important. The
fix is justified (it targets the correct optimism $b_n+S(K)$) and empirically
well-calibrated for heterogeneous candidates; at \emph{exact} ties the ordinary $n$-out-of-$n$
bootstrap of an argmax is delicate \citep{Andrews2000, FangSantos2019}.
Empirically selection-aware remained well-calibrated in the exchangeable
regime, consistent with the estimand being a smooth location shift rather
than the sampling law of the maximum, but a complete consistency
proof at exact ties---with the explicit $m$-out-of-$n$ subsampling
rate---is the \emph{one} open point left for future work. The
near-constant $\dstar$ is no longer merely empirical: it is a corollary of
the derived closed form, which reproduces the measured value from the
covariance constants to within their $\sim$10--20\% sampling wobble (the
$a_K/\sqrt2$ form being the variance-homogeneity idealization). A third caveat is
the estimand itself: under censoring Harrell's $C$ targets a
censoring-dependent quantity whose score-specific shift need not cancel in
$\Delta$, so the confirmatory analysis uses Uno's IPCW $C$ with a
censoring-model sensitivity (Section~\ref{setup-censoring}). Finally, our
scope is \emph{discrimination only}: a concordance difference is itself an
insensitive measure of a marker's added value \citep{Pencina2008}, and
incremental calibration and net benefit face the same
developed-versus-fixed asymmetry but are deferred to future work.

\textbf{Conclusion}: Ignoring the selection over candidate constructions
produces structurally over-confident claims of incremental value, and a
small, adoptable change to existing bootstrap pipelines restores valid inference. This is a cautionary tale, a drop-in fix, and guidance on when it
matters---not a claim to beat the state of the art.

\FloatBarrier% flush every remaining float so all end-matter follows the last table/figure
\backmatter

\section*{Acknowledgements}

This research was supported in part through computational resources from
Advanced Research Computing at the University of British Columbia. During
this work, the author used AI-based tools (large language models) to assist
with analysis and simulation code, text editing, and checking derivations;
the author verified all outputs and takes full responsibility for the
content.\vspace*{-8pt}

\section*{Funding and Conflicts of Interest}

This work received no specific grant funding. The author declares no
conflicts of interest.\vspace*{-8pt}

\section*{Data Availability}

The NHANES data are publicly available from the U.S. National Center for
Health Statistics (\url{https://www.cdc.gov/nchs/nhanes/}); the linked
mortality files are public-use \citep{NCHS_LMF}. Code for the method,
simulation, and NHANES experiment reproduces every result in this paper and
is available at \url{https://github.com/ehsanx/selection-aware-cindex}.\vspace*{-8pt}

\section*{Supplementary Material}

Web Appendices A--D, with full statements and proofs, the simulation
data-generating process and complete results, and the NHANES experiment
design, are provided in the Supplementary Material accompanying this
preprint.\vspace*{-8pt}

\clearpage
\bibliographystyle{abbrvnat}
\bibliography{ref}

\label{lastpage}

\end{document}